\documentclass{aa}
\usepackage{times}
\usepackage{t1enc}
\usepackage{psfig}
\usepackage{epsfig}

\begin{document} 


\title{First AU-scale observations of V1647 Ori with 
     VLTI/MIDI\thanks{Based on observations made with the 
     Very Large Telescope Interferometer at Paranal Observatory}}
\author{P. \'Abrah\'am\inst{1} 
        \and 
        L. Mosoni\inst{1}
        \and
        Th. Henning\inst{2}        
        \and 
        \'A. K\'osp\'al\inst{1}
        \and 
        Ch. Leinert\inst{2}
	\and
	S.P. Quanz\inst{2}
        \and 
        Th. Ratzka\inst{2}
        }
\offprints{P. \'Abrah\'am, email: abraham@konkoly.hu}
\institute{Konkoly Observatory of the Hungarian Academy of Sciences,
           P.O. Box 67, H-1525 Budapest, Hungary 
      \and Max--Planck--Institut f{\"u}r Astronomie,
           K{\"o}nigstuhl \/17, \/D-69117 Heidelberg, Germany
           }
\date{Received date; accepted date} 
\authorrunning{P. \'Abrah\'am et al.}  
\titlerunning{VLTI/MIDI observations of V1647 Ori}
\abstract{The young eruptive star V1647\,Ori was observed with MIDI,
the mid-infrared interferometric instrument at the Very Large
Telescope Interferometer (VLTI), on March 2, 2005.  We present the
first spectrally resolved interferometric visibility points for this
object. Our results show that (1) the mid-infrared emitting region is
extended, having a size of $\approx$7\,AU at 10\,$\mu$m; (2) no
signatures of a close companion can be seen; (3) the $8-13\,\mu$m
spectrum exhibits no obvious spectral features. Comparison with
similar observations of Herbig Ae stars suggests that V1647\,Ori
probably possesses a disk of moderate flaring. A simple disk model
with $T{\sim}r^{-0.53}, {\Sigma}{\sim}r^{-1.5},
M_{d}=0.05\,\rm{M}_{\odot}$ is able to fit both the spectral energy
distribution and the observed visibility values simultaneously.
 } \maketitle


\section{Introduction} 
\label{sect:Intro} 

In January 2004 a new reflection nebula (McNeil's Nebula) appeared in
the LDN\,1640 dark cloud of the Orion\,B molecular cloud complex
(\cite{McNeil}).  V1647\,Ori, whose outburst ($\sim$4\,mag in
the I-band) caused the appearance of McNeil's Nebula, is a low-mass
pre-main sequence object (\cite{Briceno}, \cite{abraham04a}). Its
eruptive behaviour suggests that V1647\,Ori is either an FU Orionis
(FUor) or an EX Lupi (EXor) type object, or maybe an intermediate-type
object between FUors and EXors (\cite{Muzerolle},
\cite{kospal}). Near-infrared colour maps show that the source is
embedded in an elongated disk-like structure, whose size is
approximately 7000 AU, and its inclination (the angle between the
normal of the disk and the line of sight) is about 60$^{\circ}$
(Acosta-Pulido et al., in prep.). The object had been gradually
fading until October 2005, when the eruption rapidly ended
(\cite{kospal}).

Both FUors and EXors are low-mass pre-main sequence stars which
exhibit optical brightening of several magnitudes. FUors are
characterised by outbursts of several decades, while EXors exhibit
repetitive outbursts on monthly timescale. The eruptive mechanism of
both types is thought to be a rapid temporal increase of the disk
accretion rate (\cite{HK96} and references therein). In this model the
young star is accreting a substantial amount of material from the
parent molecular cloud core ($10^{-6}$ M$_{\odot}$/yr) via a
circumstellar disk. The infalling matter piles up in the inner disk
until its surface density -- and thus opacity -- becomes high enough
to switch on a thermal instability leading to the dramatically
increased accretion rate ($10^{-4}$ -- $10^{-3}$ M$_{\odot}$/yr). It
is still debated whether perturbation due to a close companion is
needed to trigger the outburst, and it is also an open question
whether all eruptive stars have companions (e.g. \cite{wang04},
\cite{RA04b}).

The geometrical structure of the inner part of the circumstellar
material in eruptive systems -- though it is probably closely related
to the outburst mechanism -- is still unclear. Most models assume that
apart from the circumstellar disk, there is an infalling remnant
envelope of the molecular cloud core, whose inner radius is of the
order of a few AU (\cite{KH91}, \cite{Turner}). This envelope
supplies the disk with the material needed for the strong
eruptions. In the case of V1647\,Ori, Muzerolle et al. (2005) proposed
that an optically thin envelope would be necessary to explain the flat
infrared spectral energy distribution. However, a direct detection of
such envelopes around eruptive stars has not been attempted so far.

In this paper we report on AU-scale observations of V1647\,Ori at
10\,$\mu$m with MIDI, the mid-infrared interferometric instrument
mounted at the Very Large Telescope Interferometer (VLTI) of ESO's
Paranal Observatory (\cite{Leinert03}).  We present spectrally
resolved interferometric visibility points, and analyse the data with
special emphasis on the structure of the circumstellar material and on
the signature of a nearby companion.

\section{Observations and data reduction}
\label{sec:obs}

V1647 Ori was succesfully observed with MIDI on the UT3-UT4 baseline
of the VLTI on March 2, 2005. The projected baseline length was 56\,m
with a PA=112$^{\circ}$. Due to the lack of adequate guide star, 
MACAO\footnote{http://www.eso.org/projects/aot/macao\_vlti/, the
Curvature AO system for the VLTI}, could not support the observation,
which was thus affected by the seeing (0.9-1.0$''$ at the time of the
observation). Single telescope N-band spectra were also taken with
both UT3 and UT4. Additional N-band spectra were obtained on December
31, 2004 (then the interferometric part of the observation failed). On
March 2, HD\,37160 was observed as a calibrator, while on December 31
observations of HD\,50778, performed 2 hours later, were used.

The obtained data set consists of acquisition images with the N8.7
filter, 8--13\,$\mu$m low resolution spectra (R=30), and
interferometric measurements.  In the data reduction we followed the
general processing scheme as described by Leinert et al. (2004). The
measured raw visibilities of V1647\,Ori were divided by raw
visibilities obtained from the HD\,37160 data.  The observations were
reduced in two independent ways: with the MIDI Interactive
Analysis (MIA) package, which uses the power spectrum method, and the
Expert Work Station (EWS) package, which is based on a coherent,
linear averaging method. For the extraction of the spectra the two
packages\footnote{http://www.mpia-hd.mpg.de/MIDISOFT/} consider fitted
and fix masks, respectively. The interferometric results were then
compared and showed agreement better than 10\%.

All available calibrator data obtained on March 2, 2005 were
reduced in order to determine the errors of the spectrally resolved
visibilities.  After investigating the data we decided to apply a
uniform multiplicative error of 10\% as a conservative estimate.

At both epochs the N-band spectra obtained by the two telescopes
differ in their absolute flux levels by a factor of $\approx$1.4, and
their spectral shapes slightly deviate at ${\lambda}{>}12\,\mu$m.  The
reason of this artifact is not yet clarified, thus we evaluated the
acquisition images and extracted flux density values at $8.7\,\mu$m
for V1647\,Ori (for the calibrator, HD\,37160, 11.85 Jy was
assumed). Photometry taken with UT3 turned out to be more reliable,
judged from the repeatibility of the standard star measurements, and
we estimate its absolute accuracy to be 15\%. The absolute calibration
was then carried out by scaling the N-band spectra to the photometric
points at $8.7\,\mu$m.

\section{Results} 
\label{sect:results}

\begin{figure}  
\begin{center} 
\psfig{figure=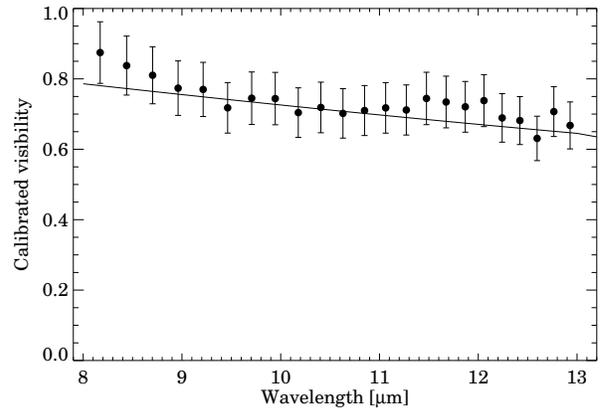,width=80mm}
\caption{Calibrated visibilities as a function of wavelength. The
uniform error bars of 10\% reflect our conservative estimate of the
uncertainties. The solid line represents the model visibility curve
of V1647 Ori (see Sect 4.).}
\label{fig:fig1}
\end{center} 
\end{figure}

Figure\,\ref{fig:fig1} shows the calibrated visibilities as a function of
wavelength.  The values gradually decrease from 0.87 to 0.75 between 8
and 10\,$\mu$m, and are approximately constant at longer wavelengths,
with a small dip between 12 and 13\,$\mu$m. This pattern clearly
demonstrates that the emitting region is resolved, and suggests a
non-uniform temperature distribution of the emitting material
(otherwise the visibilities would monotonically increase towards
longer wavelengths due to the lowering spatial resolution).

Adopting Eq.~1 from Leinert et al.~(2004), we calculated visibility
values for a series of extended model sources of increasing sizes, 
represented by Gaussian brightness profiles, and compared them with 
our observations at
different wavelengths. The derived FWHMs increase towards longer
wavelengths from 6$^{+2}_{-3}$\,mas at 8.2\,$\mu$m to 16$\pm$3\,mas at
13\,$\mu$m. At the distance of V1647\,Ori (450 pc) these angular sizes
approximately correspond to 2.7 and 7.2\,AU linear sizes,
respectively. The increasing sizes of Gaussians with increasing
wavelength indicate a radially decreasing temperature profile.

\begin{figure} 
\begin{center} 
\psfig{figure=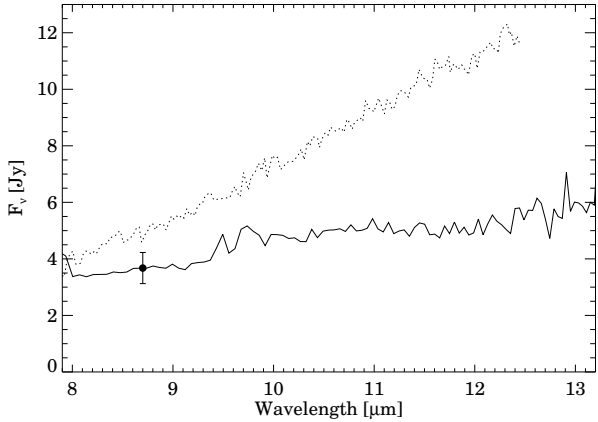,width=80mm}
\caption{N-band spectra of V1647 Ori at different epochs: {\it solid
line} -- VLTI/MIDI (March 2, 2005, this study); {\it dotted line} --
UKIRT/Michelle (March 11, 2004, Andrews et al.~2004). The MIDI
spectrum was scaled to the photometric point at $8.7\,\mu$m obtained
with UT3 ({\it filled dot}). Neither the MIDI nor the UKIRT spectrum
shows any obvious spectral features.  }
\label{fig:fig3}
\end{center} 
\end{figure}

Fig.\,\ref{fig:fig3} presents our 8--13\,$\mu$m spectrum of V1647\,Ori
obtained with MIDI on UT3 in March 2005. The spectrum is smoothly
rising towards longer wavelengths, and no obvious spectral features
(silicates, PAHs, molecular ices) can be seen. The shape of the
December 2004 MIDI spectrum is identical to the plotted spectrum
within the measurement uncertainties. For comparison, in
Fig.~\ref{fig:fig3} we also overplotted an earlier spectrum of
V1647\,Ori from March 2004 (see figure caption).  Comparison of the
spectra suggests that the N-band spectrum changed dramatically
(especially at longer wavelengths) between March and December 2004,
but exhibited an approximately constant spectral shape afterwards,
between December 2004 and March 2005.

We calculated the spectrum of the correlated flux as the product of
the measured N-band spectrum and the spectrally resolved visibilities,
in order to study the emission of the innermost part of the
circumstellar structure. We found that the main fraction of the
observed N-band flux ($\approx 70\% $) is emitted in the inner
regions. The relatively flat shape of the correlated spectrum
indicates that the innermost part has higher temperature. Like in the
full spectrum, no spectral features can be observed in the innermost
spectrum.


\section{Discussion}

V1647\,Ori is a low-mass object (\cite{Briceno}, \cite{abraham04a}).
Nevertheless, due to its increased outburst luminosity
($L_{bol}\,{=}\,44 L_{\odot}$, \cite{Muzerolle}), it may look similar
to intermediate-mass young stars. As the first step of our
analysis, we compare our MIDI results with similar observations of a
sample of Herbig Ae/Be stars (\cite{Leinert04}). Looking first at the
visibility curves (Fig.\,\ref{fig:fig1} of the present paper; Fig.\,4
in Leinert et al.~2004) one notices a general agreement between the
shapes of the V1647\,Ori and the Herbig Ae/Be curves: a faster drop in
the $8-9\,\mu$m range is followed by almost constant visibility values
at longer wavelengths (exceptions are HD\,179218 and 51\,Oph).  All
Herbig Ae/Be stars show lower visibility values than V1647\,Ori,
indicating that these objects are more resolved. This finding can be
explained to a large extent by distance differences: these Herbig
Ae/Be stars are at a distance of 103 -- 244\,pc compared with the
distance of V1647\,Ori of 450\,pc. We tested this hypothesis with the
help of a disk model (see below) in which the distance was scaled from
450 to 150\,pc, and found visibility curves similar to those of the
Herbig Ae/Be stars.

\begin{figure}
\begin{center} 
\psfig{figure=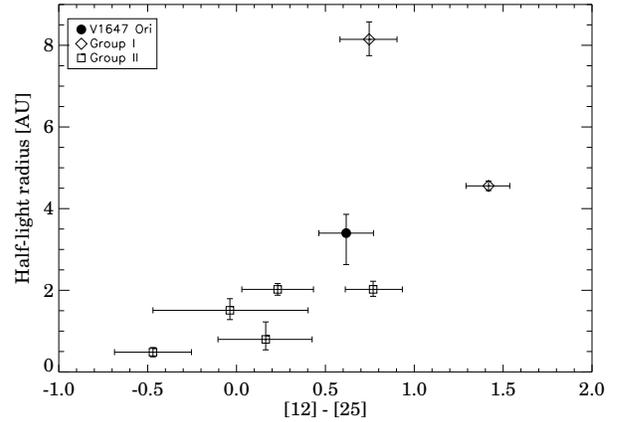,width=80mm}
\caption{Correlation between the mid-infrared spectral shape, calculated as
$[12]-[25]={-2.5\log(F_{\nu}(12\,\mu\rm{m})/F_{\nu}(25\,\mu\rm{m}))}$,
and the half-light radius (adapted from Leinert et al. 2004, see text). 
V1647\,Ori is marked with a filled circle.}
\label{fig:fig5}
\end{center} 
\end{figure}

In Fig.\,\ref{fig:fig5} we reproduced Fig.\,5 of Leinert et
al. (2004). In order to overplot V1647\,Ori in the diagram, we
computed the half-light radius as defined in Leinert et al. (2004),
and estimated the $[12]-[25]$ magnitude difference by interpolating
among photometric points measured by the Spitzer Space Telescope in
March 2004 (\cite{Muzerolle}).  When plotting the Herbig Ae/Be stars,
different symbols were used for Group\,I sources (thought to harbour
flared disks, \cite{meeus}), and Group\,II sources (believed to
possess non-flaring disks). V1647\,Ori fits into the trend that redder
disks are more extended. Considering its position between Group\,I and
Group\,II sources one may conclude that V1647\,Ori probably has a disk
of moderate flaring. However, more complex models including
additional components, such as envelopes, cannot be excluded.

FU\,Orionis, the prototype of the FUor class, was also observed with
MIDI (\cite{Quanz}), and -- being located at the same distance -- it
is natural to compare FU Ori and V1647\,Ori.  The spectrally resolved
visibility curves of FU\,Ori are very similar to that of V1647\,Ori
(at least for 2 of the 3 measured baselines) in both wavelength
dependence and absolute value. Their N-band spectra are, however,
somewhat different: FU\,Ori exhibit weak silicate emission, while
V1647\,Ori shows only smooth continuum radiation. Malbet et al. (2005)
modelled FU\,Ori with a geometrically thin accretion disk which could
reproduce their near-infrared interferometric results as well as the
broad infrared--millimetre SED. The same model, however, failed to
match the mid-infrared visibility curve obtained by MIDI
(\cite{Quanz}).

\begin{figure}
\begin{center} 
\psfig{figure=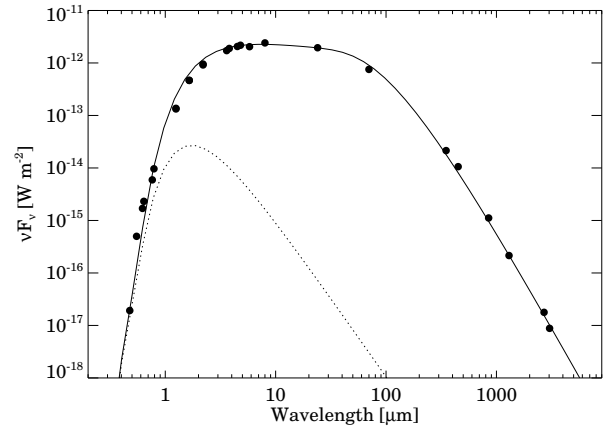, width=80mm}
\caption{Spectral energy distribution of V1647 Ori. All optical and
infrared data plotted were obtained in February-March 2004.  The thick
line is the best fitting disk model (see text); dotted line represents
contribution from the F0V star. Sources of data: {\it VRI and JHK:}
Acosta-Pulido et al., in prep.; {\it g$'$r$'$i$'$ and JHK:} Reipurth
\& Aspin 2004a; {\it 3.6--70\,$\mu$m:} Muzerolle et al. 2005; {\it
3.8, 4.8\,$\mu$m:} Vacca et al. 2004; {\it submm and mm-wavelength:}
Andrews et al. (2004), Lis et al.~(1999), Mitchell et al.~(2001),
Tsukagoshi et al. 2005.}
\label{fig:fig6}
\end{center} 
\end{figure}

In order to model the spectrally resolved visibilities of V1647\,Ori,
first we constructed a spectral energy distribution. Fortunately, many
observations were performed by different groups in February-March
2004, which cover the whole $0.5-850\,\mu$m range and are now
available in the literature. Though this date was one year earlier
than our MIDI observation, V1647\,Ori was in the outburst phase
at both epochs. The SED is plotted in Fig.\,\ref{fig:fig6} (for
references of individual observations see figure caption). The SED is
characterised by a flat spectral shape (${\nu}F_{\nu}{\approx}$const.)
in the $3-30\,\mu$m range, similarly to a number of FU Orionis
objects, like V1057\,Cyg, V346\,Nor, Z\,CMa (\cite{abraham04b,
keck}). On the other hand, the SED of V1647\,Ori significantly differs
from that of FU\,Ori (\cite{malbet}).

We model the SED using a simple disk model. More complex
configurations, e.g.~flat disk plus envelope, might also be plausible,
but the lack of any silicate emission in the N-band spectrum
(Fig.~\ref{fig:fig3}) argues against the presence of an optically thin
envelope (e.g.~like the one proposed by \cite{Muzerolle}). Our
analysis tests the adequacy of simple disk models to simultaneously
fit the SED and the MIDI visibilities. In our model one can specify
the radial profiles of the temperature and the surface density in the
disk, assuming that both follow a single power-law. We assumed
surface density as ${\Sigma}{\sim}r^{-1.5}$, an extinction of $A_V=10$
mag, and an inclination angle of 60$^{\circ}$ (Acosta-Pulido et al.,
in prep.). The central object was an F0V star, but its detailed
properties did not affect considerably the fit.  The fitted model
parameters for the disk were the following: temperature
$T(1\,\rm{AU})=680\,\rm{K}$ and $T{\sim}r^{-0.53}$, inner and outer
disk radii $7R_{\odot}$ and $100\,\rm{AU}$, respectively, disk mass
$M_{d}=0.05\,\rm{M}_{\odot}$.  The best fit to the spectral energy
distribution is overplotted in Fig.\,\ref{fig:fig6}.  The simple disk
model matches all data from 0.5 to 3000\,$\mu$m with high precision.

We predicted visibilities from the above disk model according to the
van Cittert-Zernicke theorem, i.e.~the complex visibility  is
computed as the Fourier transform of the source brightness
distribution, and vice versa. The resulting spectrally resolved
visibility curve for the exact projected baseline of the V1647\,Ori
observation is overplotted in Fig.\,\ref{fig:fig1}. The prediction
agrees with the observed visibility points within the measurement
uncertainties. Thus our simple disk model of V1647\,Ori fits the
spectral energy distribution and matches the visibility curve
simultaneously. The model is also able to reproduce the slope of our
N-band spectrum. We note, that the canonical model usually
proposed for FUors (an optically thick, geometrically thin accretion
disk with $T{\sim}r^{-0.75}$ (\cite{shakura73}), or a flared disk with
$T{\sim}r^{-0.75}$ in the inner disk and a shallower temperature
profile in the outer regions) does not fit the data. However, our
model with a shallower temperature profile of $T{\sim}r^{-0.53}$,
whose exponent is similar to that of flared disk regions in standard
models, can successfully reproduce the observations. Also, based on
recent near-infrared interferometric measurements, Millan-Gabet et
al.~(2005) found that many FUors require a different model than the
canonical one. An exception is FU Orionis itself, which was
successfully modelled by Malbet et al.~(2005) using a geometrically
thin accretion disk.

Temporal evolution of the N-band spectrum of V1647\,Ori
(Fig.~\ref{fig:fig3}) may suggest a more complex picture of the
circumstellar structure. The different fading rates at 8 and
13\,$\mu$m might indicate the presence of two physically separate
components, which have different energy budget and cooling
rate. Detailed time-dependent models will be necessary to describe
simultaneouly the SED, the visibilities and the temporal evolution.

A companion would cause sinusoidal variations in the spectrally
resolved visibilities with appropriate baseline position angles. The
shape of our visibility curve suggests that no companion is present at
the measured position angle whose separation is less than 100\,AU and
brightness ratio is greater than 10\%. Nor do the acquisition images
show any companion. The presence of companions is an important issue
to discriminate between different hypotheses explaining the outburst
mechanism (\cite{RA04b}). Due to the lack of more MIDI
observations on baselines of different length and orientation, we can
exclude the companion only along the measured baseline. The recent
rapid fading of V1647\,Ori (\cite{kospal}) may prevent us from
obtaining more visibility data in the present outburst period. A
future brightening might offer the possibility to clarify the
existence of a companion of brightness somewhat similar to the
primary.

\begin{acknowledgements}
  The work was supported by the grants OTKA T\,034584 and K\,62304 of
  the Hungarian Scientific Research Fund, and by the European
  Interferometry Initiative (Fizeau Programme). The authors thank
  Frank Przygodda for the help in the preparation of the MIDI
  observations, the ESO VLTI team (Garching/Paranal) for preparing and
  performing the service observations, and Attila Mo\'or for his work
  on the absolute calibration of the MIDI spectra. The authors are
  also grateful to an anonymous referee, whose helpful comments
  improved the paper.
\end{acknowledgements} 


\end{document}